\documentclass[prints]{aa}
\usepackage{psfig}
\begin{document}

\thesaurus{11.04.2; Galaxies: dwarf
	   11.09.1; M81 Group
	   11.16.1; Galaxies: photometry
}
\bigskip
\title{Dwarf Spheroidal Galaxies in the M81 Group Imaged with WFPC2
\thanks{Based on observations made with the
NASA/ESA Hubble Space
Telescope.  The Space Telescope Science Institute is operated by the
Association of Universities for Research in Astronomy, Inc. under NASA
contract NAS 5-26555.}}

\titlerunning{Dwarf spheroidal galaxies in the M81 group}

\author{I.D.Karachentsev \inst{1} \and V.E.Karachentseva \inst{2}
\and A.E.Dolphin \inst{3}
\and D.Geisler \inst{4}
\and E.K.Grebel \inst{5,6}\thanks{Hubble Fellow}
\and P.Guhathakurta \inst{7}\thanks{Alfred P.\ Sloan Research Fellow}
\and P.W.Hodge \inst{5}
\and A.Sarajedini \inst{8}
\and P.Seitzer \inst{9}
\and M.E.Sharina \inst{1}}

\authorrunning{I.D.Karachentsev et al.}

\institute{Special Astrophysical Observatory, Russian Academy
   of Sciences, N.Arkhyz, KChR, 357147, Russia,
\and Astronomical observatory of Kiev University, 04053, Observatorna 3, Kiev, Ukraine
\and Kitt Peak National Observatory, National Optical Astronomy Observatories, P.O. Box 26732, Tucson, AZ 85726, USA
\and Departmento de Fisica, Grupo de Astronomia, Universidad de Concepcion, Casilla 160-C, Concepcion, Chile
\and Department of Astronomy, University of Washington, Box 351580, Seattle, WA 98195, USA
\and Max-Planck-Institut f\"ur Astronomie, K\"onigstuhl 17, D-69117 Heidelbergh, Germany
\and UCO/Lick Observatory, University of California at Santa Cruz, Santa Cruz, CA 95064, USA
\and Astronomy Department, Wesleyan University, Middletown, CT 06459, USA
\and Department of Astronomy, University of Michigan, 830 Dennison Building, Ann Arbor, MI 48109, USA}

\date{Received:16 May 2000}

\maketitle
\begin{abstract}

  We obtained HST/WFPC2 images of the dwarf spheroidal (dSph) galaxies
K61, K63, K64, DDO78, BK6N, and kk77 in the M81 group. Our color-magnitude
diagrams show red giant branches with tips (TRGB) falling within
the range of $I =
[23.8 -24.0]$ mag. The derived true distance moduli (DM) of the 6 dSphs
ranging from 27.71 to 27.93 mag are consistent with their membership in the
group. Given accurate distances of 5 other group members, which have been
derived via TRGB or cepheids, the mean DM of the M81 group is $(27.84\pm0.05)$ mag.
We find the difference of the mean distances to the M81 and NGC~2403 groups to
be $D_{\rm M81} - D_{\rm NGC2403} =(0.5\pm0.2)$ Mpc, which
yields a deprojected separation of 0.9 Mpc.
With respect to the Local group, M81 and NGC~2403 have radial velocities  of
106 and 267 km~s$^{-1}$   respectively, while the velocities of the group centroids are
142 and 281 km~s$^{-1}$. The higher velocity of the closer system may indicate that
these groups are moving towards each other, similar to the Milky Way
and M31 in the Local group. Several globular cluster candidates have been
identified in the galaxies.
\end{abstract}

\section{Introduction}

  Until recently most studies of dwarf spheroidal galaxies have been carried
out in the Local Group only. The properties of the Local Group sample have
been reviewed by  Grebel (1997), Da Costa (1998), and Mateo (1998). But now
progress is being made in the study of the dwarf population of other nearby
groups, such as the M81 group, thanks to the efforts of various authors.
Being among the faintest galaxies and of very low surface brightness, dSphs
are rather difficult to recognize optically
in the sky. As gas-poor galaxies, they are usually
undetectable in the  H{\sc I} line as well. In the case of the M81 group
an additional observational difficulty arises, because of the group location
in an area of the sky that is contaminated with  dense Galactic  H{\sc I}
emission (Appleton et al. 1993)
and Galactic cirrus (Sandage 1976). Therefore, the only way
to establish the true dwarf spheroidal membership in the M81 group is by
measurement of their distances via photometry of their stellar populations.
Below we briefly outline the history of discovery of dwarf spheroidal
galaxies in the M81 group. 

  Van den Bergh (1959) published the DDO catalog of low surface brightness
objects with angular diameters $a> 1\arcmin$ revealed by visual inspection of
the POSS-I plates. In the vicinity of M81 he found a concentration of
LSB galaxies, which he classified as iregulars and spheroidals. As a prototype
of dSphs, the LG dwarf spheroidal system in Draco was chosen. Out of
13 supposed dwarfs in the vicinity of M81, the objects DDO~44, 71, 78,
and 87 were classified as dSphs. Later van den Bergh (1966) ascribed them
to luminosity class V, without morphological type indication. Later DDO 87
was detected in  H{\sc I} by Fisher and Tully (1975) and reclassified
as dIrr.

  Karachentseva (1968) made an independent visual all-sky survey of the
POSS-I prints for probable Sculptor-type
\begin{figure*}[hbt]
 \caption{Digital sky survey images of eight dwarf spheroidal members of
    M81 group. The field size is 6$\arcmin$, North is up, and East is to the left.}
\end{figure*}
\begin{figure*}
 \caption{WFPC2 images of six dSph galaxies: kk077, K61, DDO~71, K64, DDO~78,
    and BK6N produced by combining the two 600 s exposures through the
    F606W and F814W filters. Each galaxy is centered in the WF3 chip
    (WF3-FIX mode). The arrows point to the North and the East. Globular
    cluster candidates are indicated by circles.}
\end{figure*}
\begin{figure*}[hbt]
 \caption{ $R$ images of K61 (a) and K64 (b) obtained with the 6-m telescope
    with exposures of 1200 s. Both the images have about the same orientation
    (top is NE).}
\end{figure*}

\begin{figure*}[hbt]
%
%
\caption{Color-magnitude diagrams from the WFPC2 data for six dSph galaxies.
    The three panels for each galaxy show diagrams based on stars within
    the central (WF3) field, the "medium" field (the neighbouring halves
    of the WF2 and WF4 chips), and the outer field (remaining halves of the
    WF2 and WF4 chips). Each of these three fields covers an equal area
    of 800$\times$800 pixels. The solid lines in the left panel show the mean
    loci of the red giant branches of globular clusters with different
    metallicities, M15 ($-$2.17 dex). M2 ($-$1.58 dex), and NGC~1851 ($-$1.29 dex.)
    from left to right, based on Da Costa \& Armandroff (1990).}
\end{figure*}
%
%
%
\setcounter{figure}{4}
 \begin{figure*}[hbt]
\caption{Globular cluster candidates shown in the F606W+F814W filters from
    HST. In case of K64 the central background galaxy is presented.}
\end{figure*}
dwarf galaxies with an angular
diameter limit of 0$\farcm5$. Around M81 she found 14 objects, and some
of the largest of them
were identified with the DDO objects. New supposed dSphs were
K59, K61, and K64 ( = UGC~5442, Nilson 1973). Later Mailyan (1973)
rediscovered the objects K61 (=Mai 47), DDO~71 (=Mai 49), and K64 (=Mai
50) on the POSS-I prints.

  B\"{o}rngen and Karachentseva (1982) carried out a special photographic
survey with the 2-m Tautenburg Schmidt telescope to search for new dwarf
galaxies in the M81 group. All previously known objects were confirmed, and
some previously uncatalogued objects were found, in particular,
BK2N (dSph/dE),  BK5N (dSph), and BK6N (dSph).

  To clarify the morphology of supposed dwarf members in the M 81 group,
about 40 objects were imaged with the 6-m telescope (Karachentseva et
al. 1985, hereafter Atlas). Due to their very low surface brightness and
lack of a significant surface brightness
gradient (in contrast to dE galaxies), as well as to  the absence
of bright knots (in contrast to dIrr galaxies), the galaxies DDO~44, BK2N,
K59, K61, BK5N, DDO~71 = K63, K64, DDO~78, and BK6N were classified
as spheroidal dwarfs. Later Karachentsev~(1994) found on the POSS-II films
a new object, An 0946+6745, also classified as dSph.

 Using CCD data obtained with the Burrell Schmidt telescope at Kitt Peak
Caldwell et al. (1998) found several probable  dSph members of the M81
group. One of them, named F8D1, was confirmed as a new spheroidal
dwarf companion of M81 via HST observations.
 Recently Karachentseva \& Karachentsev (1998, hereafter kk-list) have
performed all-sky visual inspection of the POSS-II films for nearby
$(V_0 < 500$ km~s$^{-1}$) dwarf  galaxy candidates. As probable dSph members
of the M81 group, they included into the kk-list the following objects: DDO~
44 (=kk61), An0946+6745 (=kk77), K61 (=kk81), DDO~71 (=kk83), K64 (=kk85),
DDO~78 (=kk89), and BK6N (=kk91). The galaxies BK2N, and K59 were
omitted as background objects because of their high H{\sc I} velocities,
and BK5N bacause of its having
a small (0$\farcm$5) angular diameter. However, Caldwell et al.
(1998) imaged BK5N with HST and showed that it is a true dwarf
spheroidal galaxy belonging to the M 81 group. The same conclusion has
been drawn about DDO~44 based on HST observations (Karachentsev et al 1999).

  In this paper we consider the six dSph candidates that have not
been imaged so far with HST . We briefly review previous data and present
new results based on our HST observations,
which allow membership in the M81 group to be established.
Digital Sky Survey images of these galaxies are shown in Fig. 1, where
the size of each square is 6$\arcmin$. At the bottom of Fig. 1 we attached
DSS images of the two dSphs, F8D1 and BK5N, studied by Caldwell et
al. (1998).

\section{HST WFPC2 photometry}

  The six dSph galaxies were imaged with the Wide Field and
Planetary Camera 2 (WFPC2) aboard the Hubble Space Telescope over the
interval August 14 -- September 26, 1999, as part of an HST snapshot survey
(program GO 8192, PI: Seitzer) of nearby dwarf galaxy candidates from
the kk- list. The data were obtained with exposure times of 600 s in
the F606W and F814W filters. Fig. 2 shows the galaxy images
where both filters were combined. Each galaxy was centered on the WF3
chip. Two lines in the upper right corner indicate the North and East
directions. In the case of DDO~78 the WF4 chip is overlit by a
bright star.

  Before the HST observations were obtained, two objects, K61 and K64,
had been imaged
with a CCD at the 6-meter telescope. Their images in the $R$ band, each
about 3$\farcm1\times2\farcm7$ in size, are presented in Fig. 3.
 After removing cosmic ray hits we carried out point source
photometry of the frames with the DAOPHOT II package by Stetson et al.
(1990) implemented in MIDAS. The standard DAOPHOT/ALLSTAR procedure
was used for automatic star finding and then measuring stellar magnitudes
by fitting the point-spread function for each WF chip in each filter.
A total of 1500 -- 4000 stars/object were measured in both filters with
an aperture radius of 1.5 pixels. Using bright isolated stars, we
determined the aperture correction from the 1.5 pixel radius aperture
to the standard 0$\farcs$5 radius aperture size for the WFPC2 photometric
system; this correction is in the range of 0.40 -- 0.55 mag. Then we used
equations (1a), (1b), and (3) from Whitmore et al. (1999) to correct the
magnitudes for the charge- transfer efficiency loss, which depends on the
X- and Y- positions, the background counts, the brightness of the star
and the time of the observations. Transformation of the F606W and F814W
instrumental magnitudes to the standard ground- based $V,I$ system followed
the prescriptions of Table 10 ( here Holtzman et al. 1995) taking into
account different relations for blue and red stars separately. Because
we used a non-standard $V$ filter F606W instead of F555W, the resulting
$V,I$ magnitudes may contain systematic zero-point errors, which are
expected to be within 0.1 mag. Finally, objects with
goodness of fit parameters $\mid$SHARP$\mid > 0.3$, $\mid$CHI$\mid > 2$,
and $\sigma(V) >
0.2$ mag were excluded. The resulting color-magnitude diagrams (CMDs)
in $I, V - I$ are presented for the six objects in Fig. 4.
    
\section{Color-magnitude diagrams and distances}

  For each object the left panel of Fig. 4 shows the CMD for the central
WF3 field covering the main galaxy body. The middle panel represents the
CMD for the neighbouring
regions in the southern half of WF2 and the
eastern half of WF4 (the "medium" field), and the right panel comprises
stars found in the remaining outer halves of WF2 and WF4. All six
galaxy CMDs look rather similar to each other. Their stellar
population is represented
predominately by red stars. The number of
stars in each central field rises steeply at $I$ of about 23.9, which we
interpret as the tip of the red giant branch (TRGB).
 All the galaxies show a significant number of red stars with  $I < 23.9$,
which could be upper asymptotic giant branch (AGB) stars. Their presence
in these systems would not be surprising given that Caldwell et al. (1998)
found them in much higher S/N WFPC2 observations of two other M81 dwarfs:
F8D1 and BK5N. However, these AGB stars may also be due to a
crowding effect (Grillmair et al. 1996, Martinez-Delgado \& Aparicio 1997).

\begin{table*}[hbt]
\caption{Properties of dwarf spheroidal galaxies in the M81 group}
\begin{tabular}{ccccccccc} \\ \hline \hline

 Parameter & A0946+67&   K61 &  D71=K63&   K64 &   D78  &  BK6N  &  F8D1  &  BK5N  \\
	   &   kk077 &  kk081&   kk083 &  kk085&  kk089 &  kk091 &    ---   &    ---   \\
\hline
	   &         &       &         &       &        &        &        &        \\
 RA(1950.0)&  09 46 08 & 09 53 01&  10 01 18 & 10 03 09&  10 22 48&
10 31 00&  09 40 45&  10 00 45 \\
 D (1950.0)&  67 44 25 & 68 49 47&  66 48 00 & 68 04 19&  67 54 32&
66 16 00&  67 42 20&  68 29 54 \\
  $E(B-V)$  &   0.15  &  0.07 &   0.09  &  0.06 &   0.03 &   0.01 &   0.09 &   0.06  \\
    $A_v$     &   0.50  &  0.23 &   0.28  &  0.19 &   0.09 &   0.04 &   0.30 &   0.19  \\
    $A_i$     &   0.29  &  0.13 &   0.16  &  0.11 &   0.05 &   0.02 &   0.17 &   0.11  \\
 $a\times b (\arcmin)$ & $2.4\times1.8$ &$2.4\times1.4$&
$1.7\times1.5$ &$1.9\times0.9$& $2.0\times1.9$& $1.1\times0.7$& $2.2\times2.0$& $0.8\times0.6$ \\
   $V_T$     &  15.5   & 14.2  &  14.9   & 14.6  &  15.1  &  16.0  &  15.0: &  16.7   \\
 $(V-I)_T$   &   1.2   &  1.4  &   1.3   &  1.1  &   0.9  &   1.3  &  (1.1) &  (1.0)  \\
 $\mu_v(0)$   &  24.7   & 23.5  &  23.3   & 23.3  &  24.5  &  24.2  &  25.4  &  24.5   \\
  $h (\arcsec$)    &   25    &  32   &   22    &  21   &   28   &   18   &   54   &   12    \\
 $I_{(TRGB)}$   &  23.95  & 23.86 &  23.83  & 23.90 &  23.85 &  23.90 &  24.00 &  23.95  \\
  $(m-M)_0$   &  27.71  & 27.78 &  27.72  & 27.84 &  27.85 &  27.93 &  27.88 &  27.89  \\
 $(V-I)_{-3.5}$&   1.62  &  1.48 &   1.60  &  1.47 &   1.42 &   1.39 &   1.61 &   1.36  \\
 $[Fe/H]$    &  $-$1.5   & $-$1.6  &  $-$1.3   & $-$1.5  &  $-$1.6  &  $-$1.6  &  $-$1.1  &  $-$1.7   \\
 $D_{27}\;$, kpc &   2.6   &  2.6  &   1.8   &  2.0  &   2.1  &   1.2  &   2.3  &   0.8   \\
  $M_v$   & $-$12.84  &$-$13.87 & $-$13.22  &$-$13.43 & $-$12.83 & $-$11.88 & $-$13.14 & $-$11.33  \\
 $r{\rm (M81)},\;(\arcmin$)&   99    &  31   &  160    &  97   &  191   &  289   &  114   &  71     \\
 $R{\rm (M81)},\;$ kpc&  105    &  33   &  170    & 103   &  204   &  308   &  122   &  76     \\
 Type      &  dSph   & dSph/dIrr &  dSph   & dSph  &  dSph  &  dSph  &  dSph  & dSph    \\
 $F, \;$(mJy)   &  $<$5.5   & $<$3.4  &  $<$3.4   & $<$4.3  &  $<$4.9  &  $<$4.4  &   ---    & $<$3.6    \\
 $N_{gc}$     &    1:   &   1   &    1    &   0   &    1   &    1   &    1   &   0     \\
 $S_N$       &   7.4:  &  2.9  &   5.2   &   0   &   7.5  &  18.0  &   5.6  &   0     \\
& & & & & & &  & \\
\hline
& & & & & & &  & \\

\end{tabular}
\end{table*}

Either higher S/N data or a more sophisticated analysis of the current
data would be required to establish the existence and quantity of AGB stars
in these galaxies.
\begin{figure*}[hbt]
 \vbox{\includegraphics{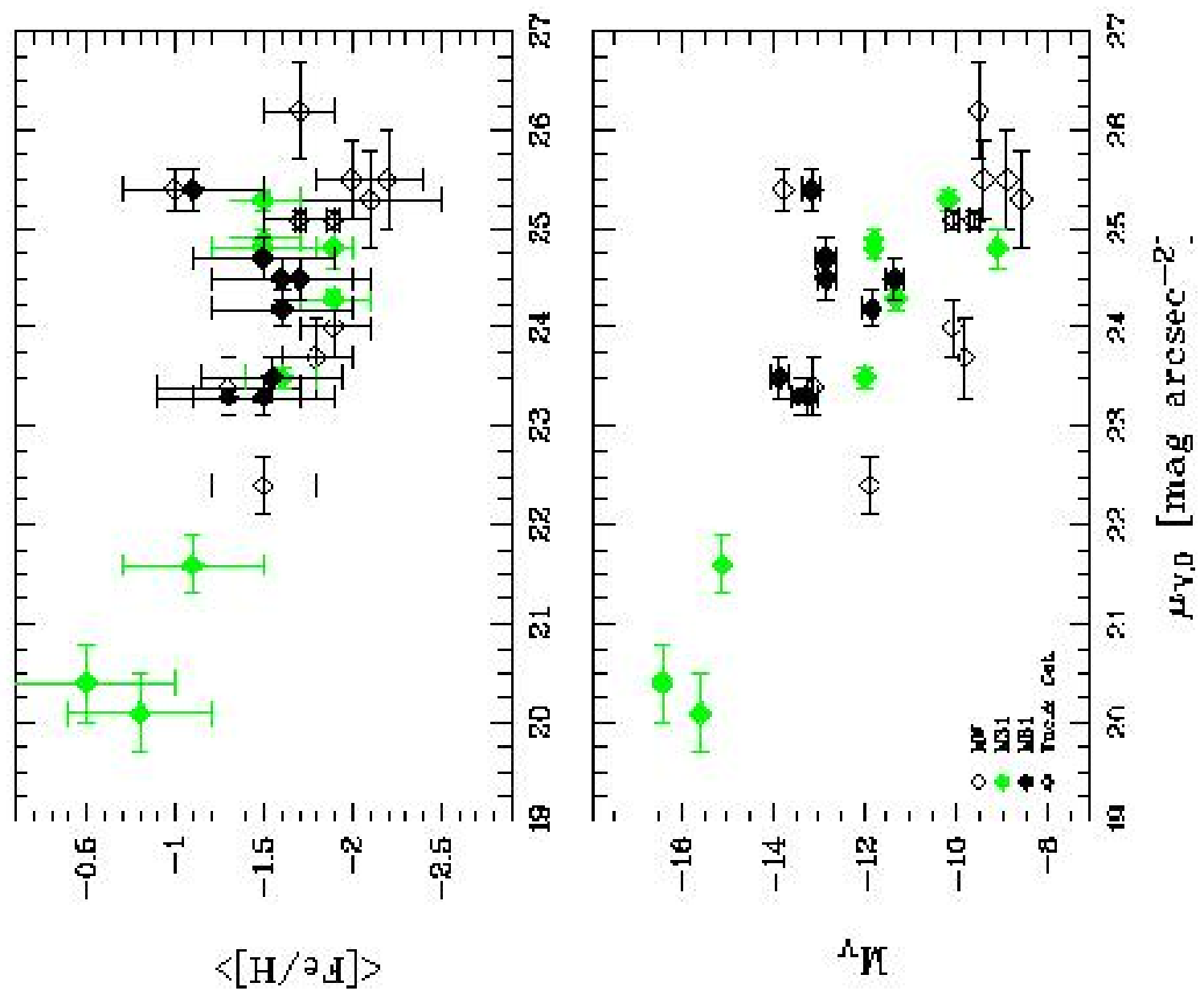}}\par

\vspace{13cm}

\caption{Mean metallicity, [Fe/H] (top), and absolute $V$ magnitude, $M_v$
    (bottom), plotted vs. central surface brightness for dwarf spheroidal
    galaxies in the Milky Way + M31 group. The M81 dSphs (dark diamonds)
    follow the same overall relationship between these global properties
    as the LG dwarfs.}
\end{figure*}

 For metal-poor systems the TRGB may be assumed to be at $M_I = - 4.05$
(Da Costa \& Armandroff 1990). We find the apparent magnitude of the
TRGB for different objects to be in the range of I(TRGB) = [ 23.83 --
23.95], which yields distance moduli in the range of [27.71 -- 27.93]
with Galactic extinctions $A_I$ from Schlegel et al. (1998). The
individual values of I(TRGB), $A_I$, and $(m-M)_0$  for the objects are
listed in Table 1. The solid lines in Fig. 4 are globular cluster
fiducials from Da Costa \& Armandroff (1990), which were reddened and
shifted to the distance of each galaxy mentioned in Table 1. The
fiducials cover the range of [Fe/H] values: $-$2.2 dex (M 15), $-$1.6 dex
 (M2), and $-$1.2 dex (NGC~1851) from left to right. The last two columns
of Table 1 refer to the objects F8D1 and BK5N studied by Caldwell et
al. (1998). After correction for Galactic extinction from the
IRAS/DIRBE results (Schlegel et al. 1998) their distance moduli, 27.88
and 27.89, lie in the same range as the others. As the one- sigma error
of the derived distance moduli we adopt a value of $\sim0\fm15$,
which includes a typical scatter in the TRGB position ($\sim0\fm1$) and an
uncertainty of zero- points ($\sim0\fm1$) under "synthetic" transformations.
    
\section{Metal abundances}
  
Given the distance moduli of these dSph galaxies, we can
estimate their mean metallicity, [Fe/H], from the mean color of the TRGB
measured at an absolute magnitude $M_I = -3.5$, as recommended by Da
Costa \& Armandroff (1990). Based on a Gaussian fit to the color
distribution of the giant stars in a corresponding $I$- magnitude
interval $(-3.5\pm0.3)$, we derived their mean colors, $(V-I)_{-3.5}$, which
lie in a range of [1.36 -- 1.48] after correction for Galactic
reddening. Following the relation of Lee et al. (1993), this provides us
with mean metallicities, [Fe/H] = [$-1.27$, $-1.61$], listed in Table 1. With
a typical statistical scatter of the mean color ($0\fm05$) and a probable
systematic error in zero points and aperture corrections ($0\fm10$) we
expect an uncertainty in metallicity to be about $\pm$0.4. Therefore
within the measurement accuracy attainable with the photometric
metallicity calibration the metallicity differences may not be significant,
but we can certainly not rule out significant overall metallicity differences
between the dSph galaxies in the M81 group.

\section{Integrated properties}

Apart from stellar photometry we also carried out aperture photometry
of each galaxy in circular apertures, which allows some global
parameters of the galaxies to be estimated: the color of the central part,
the central surface brightness, and the exponential scale length. Of
course, the accuracy of these parameters is not high because of low
surface brightness of the objects and because their extension is
comparable to the WFPC2 field size. A summary of basic properties of the
galaxies is given in Table 1. Its lines contain: (1,2) equatorial
coordinates, (3--5) Galactic extinction from Schlegel et al. (1998),
(6) galaxy dimensions along the major and minor axes approximately
corresponding to a level of $B = (26.5 - 27.0)^m/\sq\arcsec$, (7) integrated
$V$ magnitude estimated as an average of different sources, (8,9)
integrated color of the central part and the central surface
brightness from our measurements, (10) the exponential scale length in
arcsec corresponding typicaly to a distance range of [5 -- 25]$\arcsec$ from
the galaxy center, (11) apparent $I$ magnitude of the TRGB, (12)
distance modulus, (13,14) the mean color of the TRGB and corresponding 
mean metallicity, (15,16) linear diameter and the total absolute
magnitude of a galaxy, (17,18) angular and linear projected separation of
galaxy from M81, (19) morphological type, (20) an upper limit of HI
flux according Huchtmeier et al. (1999). The last two lines refer to
galaxy globular cluster candidates (see the next section). For
completeness we present also global parameters of two remaining
dSph members of the M81 group: F8D1 and BK5N from Caldwell et al. (1998)
after correction for the IRAS/DIRBE extinction. Note that these authors
estimated an apparent magnitude of F8D1 as $V_T = 13.85\pm0.25$, which
makes it the brightest dSph in the M81 group. In Table 1 we adopt for F8D1
$V_T = 15.0\pm0.5$, based on the visual appearence of the object with respect
to other dSphs.

Some additional comments about the galaxy properties are mentioned below. \\
\\
{\it kk77 = An0946+6745\/}
\\
Barely seen on the POSS-I prints because of a very low surface brightness,
the object was discovered only on the POSS-II films (Karachentsev 1994).
As with all other dSphs in the M81 group, kk77 has not been detected in
the HI line (Huchtmeier et al. 1999). \\
\\
{\em K61 = kk81 = Mai47\/}
\\
This is the brightest dSph galaxy in the M81 group and also the closest
companion to M81. After Karachentseva (1968) the object was
rediscovered by Mailyan (1973) and Bertola \& Maffei (1974). Its image
and isodensity map were presented in the Atlas (Karachentseva et al. 1985)
based on photographic observations with
the 6-m telescope. According to these data K61 has a
total magnitude $B_T$ = 15.6 and a central surface brightness $\mu_B(0) =
24.4^m/\sq\arcsec$ (Karachentseva et al. 1987). A photographic surface
photometry made on the Tautenburg plates yielded $B_T = 14.9$, $V_T =
14.3$, $\mu_B(0) = 24.4$, and an exponential scale length of $h = 33\arcsec$
(Karachentseva et al. 1984). Bremnes et al. (1998) imaged K61 with
a CCD at 1.2-m OHP telescope and derived the following parameters:
$B_T = 15.25$, $R_T = 13.69$, $\mu_B(0) = 24.7$, and $h = 32\arcsec$,
 which are in a good agreement with the
previous photographic data. Note also the results of CCD photometry
of K61 by Johnson et al. (1997): $V = 15.68$, $R = 15.10$, and
$I= 14.64$ refered to brighter (26.0 -- 26.5$^m/\sq\arcsec$) isophotes.
The maps of neutral hydrogen emission in the vicinity of M81 show the
presence of H{\sc I} in the position of K61 (van der Hulst 1979, Appleton et
 al. 1981). Van Driel et al. (1998) have made a detailed analysis of all
previous H{\sc I} observations and their own. They note difficulties in
H{\sc I} searches for K61, because of the extended H{\sc i} complex around M81
as well strong H{\sc I} emission from our Galaxy. As a result, K61 was
considered as having been undetected in the $(-260,+190)$ km~s$^{-1}$ velocity range
with a rms noise of 3.4 mJy. However, Johnson et al. (1997) revealed a bright
HII knot situated NE of the galaxy center. It shows a spectrum of
high exitation with a radial velocity of $-135\pm30$ km~s$^{-1}$. As it is seen
in Fig. 4, seven blue stars with $(V-I) < 0.4$  are observable in K61. Six of
them are located in a small ($10\arcsec\times10\arcsec$) area coincident with the HII knot.
Based on the presence of HII region and a probable presence of neutral gas,
we assume K61 to be dSph/dIrr transition type. \\
\\
{\it DDO71 = K63 = UGC5428 = Mai49 = kk83\/}
\\
Photometry of $B$ plates obtained with the 6-m telescope gives for
DDO~71 the following parameters: $B_T=15.1$ and $\mu_B(0)=24.3$. Its
luminosity profile fits well the King model (Karachentseva et al. 1987).
Bremnes at al. (1998) classified this galaxy as dE,N having an "off centre" 
nucleus.  According to the 1.2-m CCD observations,
they give $B_T = 15.95$,
$R_T = 14.27$, the central surface brightness $\mu_B(0)= 24.3$, and the
exponential scale length $h = 21\arcsec$. Huchtmeier \& Skillman (1998) detected
the galaxy in H{\sc I} with a velocity of $-126\pm5$ km~s$^{-1}$ and a line width
of $W_{50} = 27$ km~s$^{-1}$. But van Driel et al. (1998) did not detect it at a
rms noise level of 3.4 mJy; neither did Fisher \& Tully (1981) or
Schneider et al. (1992).\\
\\
{\it K64 = UGC5442 = Mai50 = kk85\/}
\\
Based on photometry obtained with the 2-m Tautenburg telescope plates
(B\"{o}rngen et al. 1982), the magnitude and
26.6$^m/\sq\arcsec$ isophote diameter are: $B=15.5$, $a=2\farcm0$,
and $V=14.7$, $a=2\farcm1$.
Later photometry by Karachentseva et al. (1984) gives $B=
15.7$ and $V = 14.8$. The same value, $B_T= 15,7$, was derived from
photometry of the 6-m telescope plates. In the Atlas the galaxy shows a
regular elliptical shape with an axial ratio of 0.5 and a low brightness
gradient, without structural complexities and knots.
The bright "nucleus" in its center was suggested as a foreground star.
\begin{table*}[hbt]
\caption{Globular cluster candidates}
\begin{tabular}{ccccccccc} \\ \hline \hline

 Parameter  &A0946+67&   K61 &  D71=K63&   K64 &   D78 &   BK6N &   F8D1 &   D44   \\
	    &  kk077 &  kk081&   kk083 &  kk085&  kk089&   kk091&   ---   &  kk061 \\
\hline
      & & & & & & & & \\

   $V_T$      & 22.83  &  20.70&   20.95 & (19.5)&  19.45&   21.80&   21.68&   22.18\\
   $V-I$      &  0.95  &   1.10&    1.11 & (1.52)&   1.11&    1.24&    0.76&  0.83\\
  $(V-I)_0$    &  0.74  &   1.00&    0.99 & (1.44)&   1.07&    1.22&    0.64&  0.77 \\
 $R(0.5L),\;(\arcsec)$&  0.45  &   0.20&    0.26 & (1.08)&   0.30&    0.29&    0.18&    0.27 \\
 $R(0.5L), \;$ pc&  8.0   &   3.6 &    4.6  &   ---  &   5.3 &    5.1 &    3.2 &     4.2  \\
 $\mu_v(0)$     & 22.4   &  18.5 &   19.2  & (19.6)&  18.0 &   20.2 &   20.2 &   20.8  \\
  $M_v$     & $-$5.51  &  $-$7.37&   $-$7.17 &   ---   &  $-$8.48&   $-$6.08& $-$6.46&   $-$5.48 \\
 Separ., kpc&  1.10  &   0.08&    0.10 & (0.12)&   0.26&    0.74&    0.25&    0.47 \\
      & & & & & & & & \\
\hline
      & & & & & & & & \\

\multicolumn{9}{l}{K64: The central diffuse object is not a globular cluster, but apparently background
galaxy.}
\end{tabular}
\end{table*}
The luminosity profile of K64 is well fit an exponential law down to the
28$^m/\sq\arcsec$ level (Binggeli \& Prugniel 1994).The CCD photometry
by Bremnes et al. (1998) gives for K64: $B_T=15.46$, and $R_T=14.04$ with
a central surface brightness and an exponential scale length,
$\mu(0)=24.2 (B)$, $23.0 (R)$, and $h= 21\arcsec(B)$, $23\arcsec(R)$,
respectively. The object was surveyed
in H{\sc I} but not detected by Schneider et al. (1992), van
Driel et al. (1998), and Huchtmeier \& Skillman (1998).\\
\\
{\it DDO78 = kk89 \/}
\\
According to Karachentseva et al. (1987), DDO~78 has a very flat surface
brightness profile with $\mu_B(0) =25.1$, $r_{ef}=32\arcsec$, and $B(t)=15.8$. It was not
detected in H{\sc I} by Fisher \& Tully (1981) and van Driel et al. (1998).
But observations with the 100-m Effelsberg radio telescope (Huchtmeier
 et al. 2000) yield an emission with $V_h = 2788\pm2$ km~s$^{-1}$ and a linewidth
$W_{50} = 32$ km~s$^{-1}$, which is probably caused by a background galaxy.\\
\\
{\it BK6N = kk91 \/}
\\
BK6N was discovered by B\"{o}rngen and Karachentseva (1982) on the
Tautenburg 2-m telescope plates. Its image and isodensity map are
presented in  the Atlas, where BK6N  has an elliptical shape with an
axial ratio of 0.5, a very low surface brightness gradient without any
structural details. Bremnes et al. (1998) presented an underexposed
image of BK6N. Photographic photometry by B\"{o}rngen et al. (1982) yields
only a rough total magnitude estimate, $B_T = 15.7$; no more optical data
were published. BK6N was not detected in the H{\sc i} line by Huchtmeier \&
Skillman (1998) and van Driel et al. (1998).

\section{Globular clusters}
The brightest dSph galaxies in the Local Group contain
globular clusters. Harris \& van den Bergh (1981) proposed to describe
their abundance by a specific frequency

$$ S_{N} = N_{gc} \cdot dex [0.4 \cdot (15 + M_{v})], $$
\noindent which is the number of observed globular clusters normalized to a
galaxy with $M_v = -15$. According to Miller et al. (1998)  nucleated
and non-nucleated dE+dSph galaxies have specific frequencies of
$6.5\pm1.2$ and $3.1\pm0.5$, respectively.

  We searched for globular clusters in our galaxies and found 5
candidates with appropriate colors and magnitudes. Their numbers per
galaxy and corresponding specific frequencies are presented in the last
two lines of Table 1. In Fig. 3 these candidates are indicated by
circles. Enlargements of their HST images are given in Fig. 5.
Table 2 presents basic photometric data for the globular
cluster candidates. Its lines give: (1,2) integrated apparent
magnitude and color, (3) integrated color after correction for Galactic
reddenning, (4,5) angular and linear half-light
radius, (6) the central
surface brightness, (7) integrated absolute magnitude, and (8) linear
projected separation of probable globular cluster from the galaxy center.
The last two column give similar parameters for another two dSphs in the
group from Caldwell et al. (1998) and Karachentsev et al. (1999). All the
globular cluster candidates have integral colors [0.64 -- 1.22], absolute
magnitudes [--5.5 -- --8.5], and half-light radii [3.2 -- 8.0] pc, which are within
\newpage
\begin{figure}[hbt]
 \vbox{\includegraphics{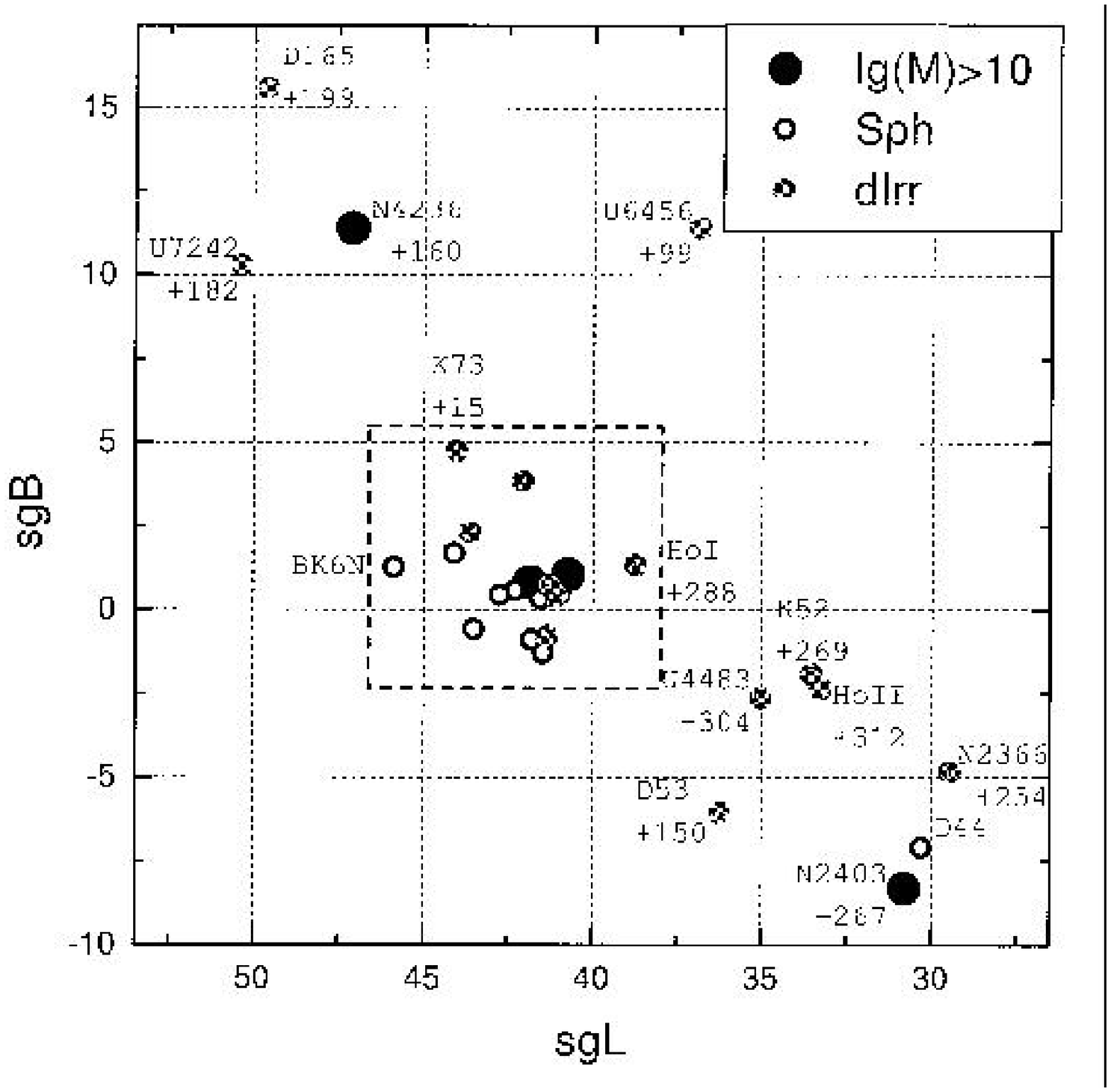}}\par

\vspace{10cm}
 \vbox{\includegraphics{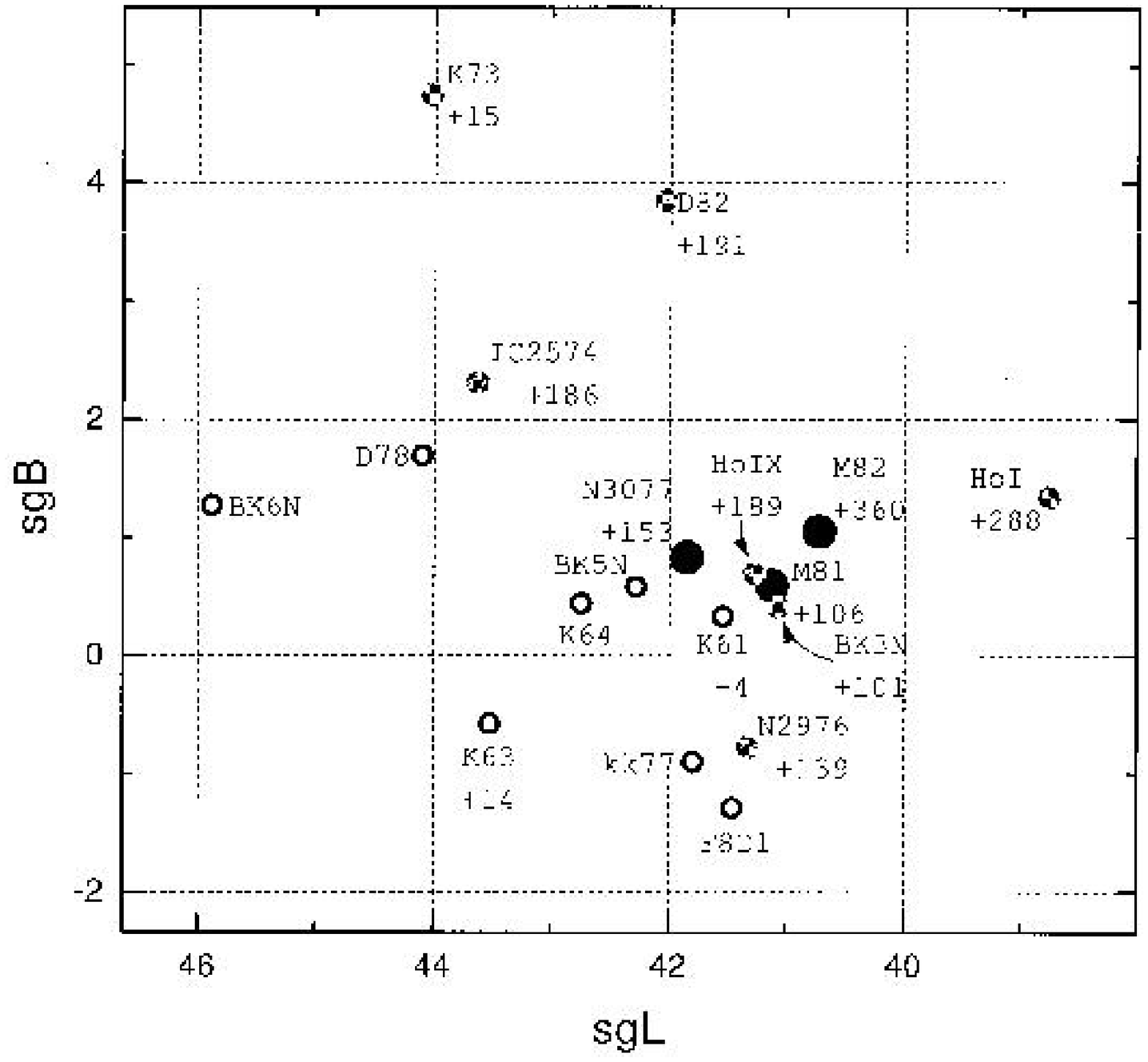}}\par
\vspace{10cm}
\caption{The distribution of galaxies in supergalactic coordinates in the
    wide vicinity of the M81 group. Top: overview of the entire region,
    bottom: the central part of the M81 group.}
\end{figure}
values typical of Galactic globular clusters. All of them,
except the kk77 candidate, show pronounced concentration towards the
galaxy centers. Considering its excentric position, as well as its low
central surface brightness, the globular cluster candidate in kk77 may be
a background galaxy. The case of K64 needs also a special
comment. Binggeli \& Prugniel (1994) considered its central star-like
object to be a "quasi-stellar nucleus". However, the large-scale HST image
of it (Fig. 5) indicates that this object is a remote red galaxy.
Therefore, both "nucleated" dSphs in M81 group: K 64 and BK5N (Caldwell
et al. 1998) are instead probable non-nucleated spheroidal systems. For the
total sample of 9 dSphs in M81 group their median specific frequency
of globular clusters, $<S_N> = 5.6$, agrees well with the data of Miller
et al. (1998).

\begin{table*}[hbt]
\caption{Projected separations and absolute magnitudes
	     of dSph companions of Milky Way, M31 and M81}
\begin{tabular}{lrr|lrr|lrr} \\ \hline \hline

\multicolumn{3}{c|}{Milky Way}&
\multicolumn{3}{c|}{M 31}&
\multicolumn{3}{|c}{M 81} \\ \hline
 Name & D, kpc&  Mv     &  Name  &D, kpc&  Mv &   Name&  D, kpc&  Mv  \\
		 \hline
      &      &       &        &     &     &       &       &      \\
 Sagittarius & 24&$-$13.8  &  AndI  & 45  &$-$11.8&   K61 &   33  &$-$13.8 \\
 Ursa Minor &  66  & $-$8.9  &  AndIII& 67  &$-$10.2&   BK5N&   76  &$-$11.3 \\
 Sculptor&  80  & $-$9.8  &  N185  & 97  &$-$15.6&   K64 &  103  &$-$13.4 \\
 Sextans &  86  & $-$9.5  &  N147  &101  &$-$15.1&   kk77&  105  &$-$12.8 \\
 Draco&  86  & $-$8.6  &  AndV  &109  & $-$9.1&   F8D1&  122  &$-$13.1 \\
 Carina  & 100  & $-$9.4  &  AndII &141  &$-$11.8&   D71 &  170  &$-$13.2 \\
Fornax & 140  &$-$13.1  &  Cassiopeia  &225  &$-$12.0&   D78 &  204  &$-$12.8 \\
 LeoII& 210  &$-$10.1  &  Pegasus &278  &$-$11.3&   BK6N&  308  &$-$11.9 \\
 LeoI & 250  &$-$11.9  &        &     &     &       &       &      \\
\hline
 Mean & 116  &$-$10.6  &        &133  &$-$12.1 &       &  140  &$-$12.8  \\
$\pm$  &  24  &  0.7  &        & 29  &  0.8&       &   30  &  0.3  \\
 $\sigma$&  68  &  1.9  &        & 78  &  2.1&       &   80  &  0.8   \\
\hline
\end{tabular}
\end{table*}

\section{Subsystem of dSphs in the M81 group}
Fig. 6 presents the distributions of dwarf spheroidal
galaxies of the Local Group according to their
absolute magnitude (lower panel) and metallicity (upper panel) versus
the central surface brightness based on the data of Grebel (2000).
Except for Sagittarius, the companions of the Milky Way (open diamonds)
and M31 (grey diamonds) follow the well-known
sequences that have been discussed by various authors (Lee 1995,
Grebel \& Guhathakurta 1999).
Here the positions of 8 spheroidal companions of M81 are shown by dark
diamonds. Except for F8D1 discussed by Caldwell et al. (1998), all dSphs of
the M81 group follow the same common relations. However, one can note that
the scatter of the M81 dSphs in the diagrams is not as great as for the
LG dwarfs.

  From the dynamical point of view it seems reasonable to consider
the companions of MW and M31 as separate subsystems. The three giant
spiral galaxies MW, M31, and M81 have the same morphological type, Sb,
and approximately the same masses, $\sim(3-4) 10^{11} M_{\sun}$, within
the standard optical radius. All known spheroidal companions of them are
presented in Table 3, where
dSphs are arranged in order of their separation from the parent galaxy.
According to these data each giant Sb galaxy has almost the same number
of dSphs: 9, 8, and 8. The average and the maximum dimension of the three
dSph subsystems do not differ significantly from one another. However, because
the companions of MW have 3D distances, but not projected ones, the dimensions
of dSph subsystems for M31 and M81 look a bit more extended. As one
can see from the third column of Table 3, the mean luminosity of spheroidal
companions of MW is much lower than for M81 companions. The
difference in absolute magnitudes of dSphs for MW and M81 is significant
at a $3\sigma$ level. Several reasons may be proposed to explain this
difference.

 a) When searching for dSphs in the M81 group, many very faint
objects were missed. The faintest known spheroidal system around M81 has $M_v = -11.3$, but
\newpage
\begin{figure}[hbt]
 \vbox{\includegraphics{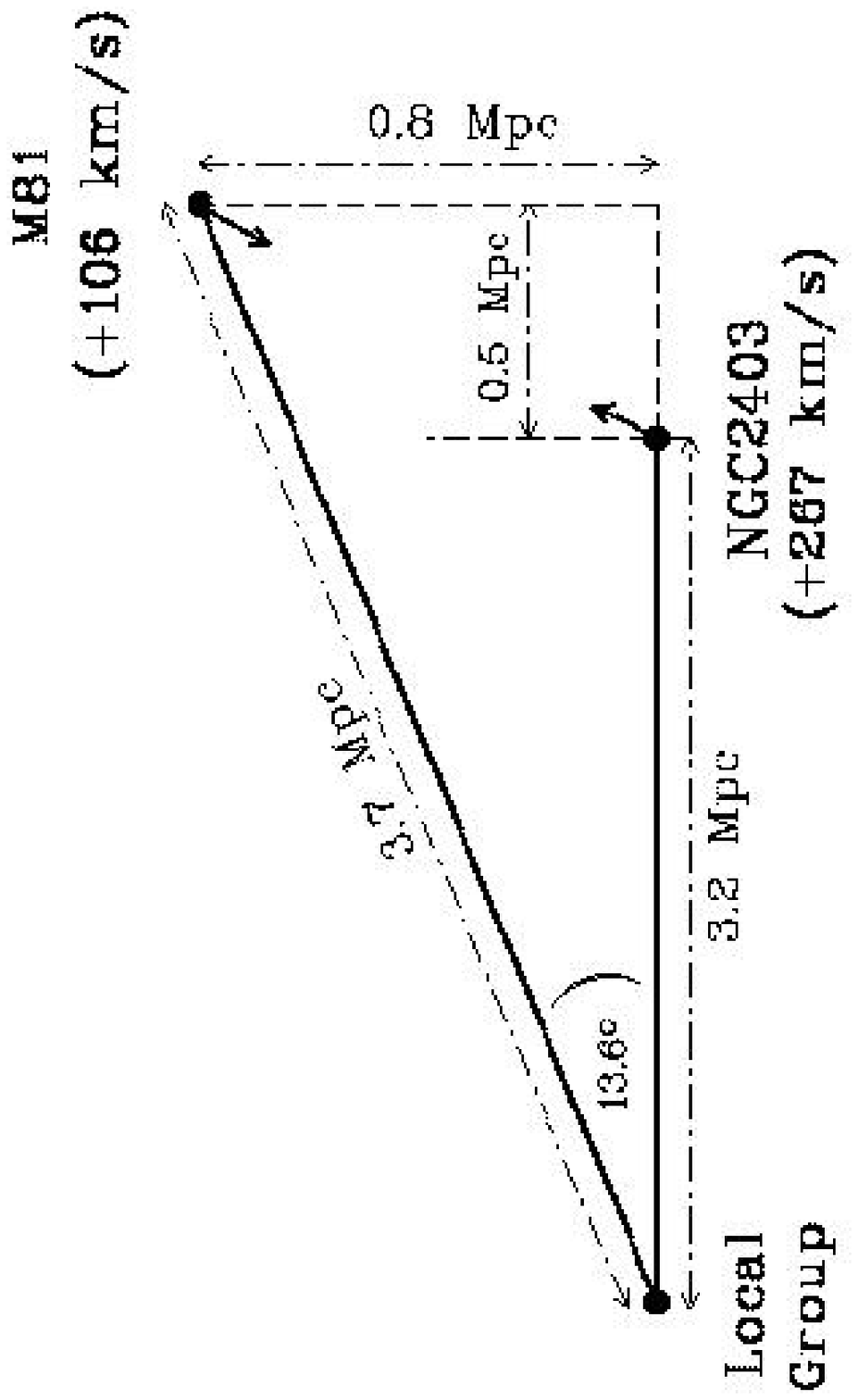}}\par
\vspace{8cm}
\caption{A schematic disposition of M81 and NGC~2403 subgroups with respect
    to the Local group.}
\end{figure}
the fraction of such faint objects among dSph companions
of MW consists of 2/3. Therefore, assuming a similar luminosity function
for both the groups leads to an expected total number of dSphs in the M81
group of $\sim$24. However, recent careful searches for new dSph objects
in the vicinity of M81 have revealed only two new probable spheroidal dwarfs
with $M_v \sim-10,-11$ mag (Karachentsev \& Karachentseva 2000) and (Froebrich
\& Meusinger 2000). b) Another cause of the discussed difference may
be systematic underestimate of the total apparent magnitudes for the most
faint and diffuse companions of MW, like Ursa Minoris, Draco, and Carina.
But the effect of underestimating must reach 2--3 mag, which seems to be
quite improbable. c) Possibly the observed difference may be related to
kinematic and dynamic features of the groups themselves such as interaction
between M81 and M82.

\begin{table*}[hbt]
\caption{Precise distance moduli for  members of the M81/NGC~2403 complex}
\begin{tabular}{lccll} \\ \hline \hline
Galaxy& $(m-M)_0\pm\sigma$& $Ai$ &Method& Source\\  \hline
M81& 27.70$\pm0.20$& 0.16& 30 cephs& Freedman et al. (1994)\\
BK5N& 27.89$\pm0.15$& 0.11& TRGB& Caldwell et al. (1998)\\
F8D1& 27.88$\pm0.10$& 0.17& TRGB& Caldwell et al. (1998)\\
UGC6456& 28.23$\pm0.10$& 0.07& TRGB& Lynds et al. (1998)\\
M82& 27.74$\pm0.14$& 0.26& TRGB& Sakai \& Madore (2000)\\
kk77& 27.71$\pm0.15$& 0.29& TRGB& present paper\\
K61& 27.78$\pm0.15$& 0.13& TRGB& present paper\\
DDO71& 27.72$\pm0.15$& 0.16& TRGB& present paper\\
K64& 27.84$\pm0.15$& 0.11& TRGB& present paper\\
DDO78& 27.85$\pm0.15$& 0.05& TRGB& present paper\\
BK6N& 27.93$\pm0.15$& 0.02& TRGB& present paper\\  \hline
Mean& 27.84$\pm0.05$& & & \\
 \hline
& & & & \\
NGC2403& 27.49$\pm0.24$& 0.08& 8 cephs& Freedman \& Madore (1988)\\
DDO44& 27.52$\pm0.15$& 0.08& TRGB& Karachentsev et al. (1999) \\
 \hline
Mean& 27.50$\pm0.02$& & & \\
 \hline
\end{tabular}
\end{table*}

\section{Distribution of dwarfs in the M81/NGC2403 complex}
The general distribution of galaxies in an extended vicinity around M81 is
shown in Fig. 7 in supergalactic coordinates. The five largest galaxies with
masses $\lg(M/M_{\sun}) > 10$ are indicated with large filled circles. Small
filled and open circles denote irregular and spheroidal dwarfs, respectively.
The numbers under the galaxy names correspond to their radial velocities in
km~s$^{-1}$ with respect to the Local group centroid. A more detailed map of the
central part of the group is given in the bottom panel of Fig. 7. According
to the group membership criterion proposed by Karachentsev (1994), dwarf
irregular galaxies NGC~2366, UGC~4483, Ho~II, and K52 are associated
with the Sc galaxy NGC~2403. Apparently DDO~44 is also one of its companions
too (Karachentsev et al. 1999). A multiple system on the opposite side
of M81 is formed by the Scd galaxy NGC~4236 and two irregular dwarfs, DDO~165
and UGC~7242. The remaining dwarf galaxies in Fig. 7 are probable companions
of M81, particularly, the most remote ones UGC~6456 (= VIIZw403) and
DDO~53.

  We do not discuss here the dynamic state of the M81 group contemplating
to do it as soon as accurate distances of all irregular galaxies in the
group are measured. We note only the distinct effect of morphological
segregation in the group: the median separations of dSphs and dIrrs from
M81 have a ratio of 4:7. The same tendency is seen also for companions of
NGC~2403 and for the Local Group.
As Fig. 7 shows, the dwarf spheroidal galaxies are distributed
around M81 asymmetrically. With respect to the group centroid (situated
between M81 and M82) all 8 dSphs are concentrated
in one quadrant. This feature may be
explained by observational selection related to the presence
of faint cirrus in the vicinity of M81 (see Fig. 4A in Bremnes et al.
1998), in which case 3/4 of the dSphs would be undetected,
and their true number in the group would be about 32.

  Estimates of distance moduli for members of the
M81/NGC2403 complex measured via TRGB or cepheids are presented in Table 4.
The original values of moduli were corrected for galactic extinction
according Schlegel et al. (1998) indicated in column (3). In all cases
when TRGB method was applied we used a value of $M_I$(TRGB) = $-$4.05.
For the six dwarf spheroidal galaxies studied here the mean distance modulus
is 27.80 with a formal error of 0.04 mag. The true error may be 2 -- 3 times
larger, taking into account errors of photometric zero -points and also
stellar crowding effect (Madore \& Freedman 1995). Nevertheless, distance
moduli for F8D1, BK5N, M82, and UGC~6456 derived via TRGB by other authors
are in reasonable agreement with our data.
 Especially good agreement is found for the distance modulus for M81 measured by Freedman et al.
(1994) via 30 cepheids. Thus, 11 members yield for the M81 group a mean
distance modulus $(m-M)_o = 27.84\pm0.05$ with a dispersion of individual
estimates of only 0.15 mag. The last value corresponds formally to the
dispersion of radial distances of the galaxies, $\sim260$ kpc, which is
comparable with the projected linear dimension of the group.

  The second subgroup of galaxies associated with NGC~2403 has only two
accurate distance estimates (the last lines in Table 4), and they show
excellent mutual agreement. Relying on these data, we conclude that there
is a systematic difference in distance moduli between M81 and NGC 2403
subgroups of $+0.34\pm0.05$ mag. This means that NGC~2403 is situated
0.54 Mpc closer to us than M81. Fig. 8 shows a scheme of their location with
respect to the Local Group. Note that the more distant galaxy M81 has a
radial velocity lower than that of NGC~2403. To explain this fact we must
assume that these galaxies have peculiar
velocities directed towards each other. The difference of their radial
velocities, $-$161 km~s$^{-1}$, seems to be rather high. Part of it is probably
caused by the motion of M81 with respect to M82. For the centroids of both
subgroups the difference is $-$139 km~s$^{-1}$ (or $-$112 km~s$^{-1}$) depending on whether
the three unreliable velocity measurements (for BK3N, K63, and K73) are
included or disregarded.

  Therefore, one might think that the two subgroups around M81 and NGC~2403
have a spatial separation of 0.94 Mpc, and are approaching  each other at
a velocity of $\sim$110 -- 160 km~s$^{-1}$.
Such a situation resembles the Local complex, where the Milky
Way and M31 subsystems are separated by $\sim0.8$ Mpc move towards each
other at 120 km~s$^{-1}$. Probably, the merger of tight binary groups is a common
phenomenon on a scale of $\sim1$ Mpc.

\acknowledgements
{Support for this work was provided by NASA through grant GO--08192.97A from
the Space Telescope Science Institute, which is operated by the Association
of Universities for Research in Astronomy, Inc., under NASA contract
NAS5--26555.  IDK, VEK, and EKG acknowledge partial support through the
Henri Chr\'etien International Research Grant administered by the American
Astronomical Society.  EKG acknowledges support by NASA through grant
HF--01108.01--98A from the Space Telescope Science Institute. This work
has been partially supported by the DFG--RFBR grant 98--02--04095.

This research has made use of the NASA/IPAC Extragalactic
Database (NED) which is operated by the Jet Propulsion
Laboratory, California
Institute of Technology, under contract with NASA. We also used
NASA's Astrophysics Data System Abstract Service and  the SIMBAD
database, operated at CDS, Strasbourg.  The Digitized Sky
Surveys were produced at the Space Telescope Science Institute under U.S.\
Government grant NAG W-2166.  We thank the referee, A.Aparicio, for
valuable comments.

{}

\end{document}